\documentstyle[12pt]{article}
\begin{document}
\begin{center}
{\large \bf INTRODUCTION TO THE POMERON} \footnote{Invited talk 
at the XXVIth Symposium on Multiparticle Dynamics, Faro, Portugal, 
September 1996}\\  
\bigskip
J. Kwieci\'nski\\
Department of Theoretical Physics\\
H. Niewodnicza\'nski Institute of Nuclear Physics\\
Krak\'ow, Poland. \\
\end{center}
\vspace{40mm}
\begin{abstract}
Elements of the pomeron  phenomenology within 
the Regge pole exchange picture are recalled.  This includes discussion 
of the  high energy behaviour of total cross-sections, the triple pomeron 
limit of the diffractive dissociation and the single particle distributions 
in the central region. The BFKL pomeron and QCD expectations for the small 
 $x$
behaviour of the deep inelastic scattering structure functions
are 
discussed.  The dedicated measurements of the hadronic final
state in deep inelastic scattering at small $x$ probing the QCD 
pomeron are described.  
The deep inelastic diffraction is also discussed.
\end{abstract}
\newpage

The term "pomeron" corresponds to the mechanism of diffractive scattering at 
high energy. It is relevant for the description of several  
phenomena and quantities like the total cross-sections $\sigma_{tot}(s)$ 
and their energy 
dependence, the real part of the scattering ampltude, 
the variation with energy of the differential elastic cross-section $d\sigma/dt$, 
behaviour of the diffractive cross-section $d\sigma/dtdM^2$, behaviour of 
the deep-inelastic scattering structure function $F_2(x,Q^2)$ at low $x$, 
behaviour of the diffractive structure function etc.\cite{DERRICK}.\\

The simplest yet presumably very incomplete description of the pomeron  
is within the Regge pole model \cite{PCOLLINS}.   
In this model one assumes  that a pomeron is described by a Regge pole 
with the  trajectory $\alpha_P(t) = 
\alpha_P(0)+\alpha^{\prime}t$. The scattering amplitude 
corresponding to the pomeron 
exchange is  given by the following formula: 
\begin{equation} 
A(s,t)= -2g(t) {exp(-i{\pi \over 2}\alpha_P(t))\over 
sin({\pi \over 2}\alpha_P(t))}\left({s\over s_0}\right)^{\alpha_P(t)}
\label{regge}
\end{equation}
where the function $g(t)$ describes the pomeron coupling. The coupling $g(t)$ 
factorizes i.e. if the amplitude $A(s,t)$ describes elastic scattering 
$a+b \rightarrow a+b$ then $g(t)=g_a(t)g_b(t)$. 
Using the optical theorem one gets the following high energy 
behaviour of the total 
cross-sections:
\begin{equation}
\sigma_{tot}(s) \sim {Im A(s,0)\over s} = {g(0)\over s} \left(
{s\over s_0}\right)^{\alpha_P(0)}
\label{sigtot}
\end{equation}
One also gets: 
\begin{equation}
\rho ={ReA(s,0)\over ImA(s,0)}=ctg({\pi \over 2}\alpha_P(0))
\label{rho}
\end{equation}
with corrections from low lying Regge trajectories which vanish at high 
energies approximately as $s^{-1/2}$.  
It follows from (\ref{sigtot}) that the Regge-pole model of a pomeron can 
describe the increase of the total 
cross-sections with energy assuming $\alpha_P(0) 
> 1$ but this parametrization will eventually violate the Froissart-Martin 
bound. \\  

Phenomenological description of $\sigma_{tot}$ and of $d\sigma/dt$ proceeds 
in general along the following two lines: 
\begin{enumerate}
\item 
One introduces the "effective" (soft) pomeron with relatively low value 
of its intercept $\alpha_P(0) \approx 1.08$ \cite{DOLACX} which 
can very well describe the high energy behaviour of all hadronic and 
photoproduction cross-sections (with possible exception of the one 
CDF point).  In phenomenological analysis one also adds the reggeon 
contribution which gives the term $\sigma_{tot}^R \sim s^{\alpha_R(0)-1}$ 
with $\alpha_R(0) \approx 0.5$. The power-like 
increase of the total cross-section has to be, of course, 
slowed down 
at asymptotic energies but  those corrections 
are presumably still relatively unimportant at presently available energies.  
The term "soft pomeron" reflects the fact that bulk of the inelastic
processes contributing to the total cross-section are the low
$p_t$ soft processes. 

\item 
One considers from the very beginning the unitarized amplitude using 
the eikonal model \cite{TTWU}.  In this model 
the partial wave amplitude   $f(s,b)$ has the form 
\begin{equation} 
f(s,b)={1\over 2i}[exp(-2\Omega(b,s))-1]
\end{equation} 
\begin{equation}
\Omega(s,b)=h(b,s)s^{\Delta}{exp(-i{\pi \over 2}\Delta)\over 
cos({\pi \over 2}\Delta)} 
\label{eikonal} 
\end{equation} 
where $\Delta >0$ and $h(b,s)$ is the slowly varying function of $s$.  
The variable $b$ denotes the impact parameter. The eikonal function can be 
viewed upon as originating from the "bare" pomeron with its intercept 
$\alpha_P(0)=1+\Delta$ being above unity. The eikonal models 
with  $\Omega(s,b) \sim s^{\Delta} (\Delta > 0)$ lead to the scattering on 
the expanding (black) disk at asymptotic energies. The radius of the disk 
grows logarithmically with increasing energy (i.e. $R(s) \sim ln(s)$ ). This  
leads to the saturation of the Froissart-Martin bound at asymptotic energies 
and to the  deviation of the shape of the  
diffractive peak from the simple exponential.  Inelastic diffractive 
scattering becomes peripheral at asymptotic energies \cite{MAOR}.  \\
\end{enumerate} 
The cross-section of the diffractive dissociation $a+b \rightarrow X +
b^{\prime}$ 
is given by the following formula: 
\begin{equation} 
s{d^2\sigma \over dt dM^2_X}={1\over 16 \pi s}
{4\over sin^2({\pi \over 2}\alpha_P(t))}g_b^2(t)
\left({s\over M_x^2}\right)^{2\alpha_P(t)} A_{Pb;Pb}(M_X^2,t)
\label{difdisab}
\end{equation} 
where, as usual, $s=(p_a+p_b)^2, M_X^2=(p_a+p_b-p_{b^{\prime}})^2$ 
and $t=(p_b-p_{b^{\prime}})^2$. The function $A_{Pb;Pb}(M_X^2,t)$ is 
the absorptive part of the forward $pomeron + b$ "scattering" amplitude.  
For large $M_X^2$ (but for $s>>M_X^2$) 
one can assume that this "scattering" 
is dominated by pomeron exchange i.e. 
\begin{equation} 
A_{Pb;Pb}(M_X^2,t)=g_{3P}(t)g_a(0) \left({M_X^2\over s_0}\right)^{\alpha_P(0)}
\label{tpa}
\end{equation}
where $g_{3P}(t)$ is the triple pomeron coupling.  
The differential diffractive 
cross-section in the triple Regge limit takes then the following 
form: 
\begin{equation} 
 s{d^2\sigma \over dt dM^2_X}={1\over 16 \pi s}
{4\over sin^2({\pi \over 2}\alpha_P(t))}g_b^2(t)
\left({s\over M_X^2}\right)^{2\alpha_P(t)} g_{3P}(t)g_a(0) 
\left({M_X^2\over s_0}\right)^{\alpha_P(0)}
\label{tpsig}
\end{equation} 
For $\alpha_P(t)=1$ this formula gives the $1/M_X^2$ spectrum for the 
diffractively produced system. 
The inelastic diffraction is affected by the multiple scattering corrections 
which lower the effective intercept of the pomeron and this effect 
is visible in Tevatron data \cite{GOULIANOS,TEVATRON}. \\  

The single particle distributions in the central region are controlled 
by the double pomeron exchange diagram which gives the rapidity plateau: 
\begin{equation}
{d\sigma\over d^2p_t dy} = f(p^2_t)\left({s\over s_0}\right)^{\alpha_P(0)-1}
\label{central}
\end{equation} 
where $y$ denotes rapidity of the produced particle.   
 The inclusive cross-section  for particle production in the central region 
 is not affected (asymptotically) by the rescattering
corrections \cite{AGK}.  This 
 implies that for the (bare) pomeron with intercept above unity  the 
 height of the plateau  should increase as 
 $\left({s\over s_0}\right)^{\alpha_P(0)-1}$ with increasing $s$.  
 Integrating the inclusive cross-section over the available phase-space 
 we get: 
 \begin{equation} 
 <n>\sigma_{tot}=const \left({s\over s_0}\right)^{\alpha_P(0)-1} ln 
 \left({s\over s_0}\right)
 \label{n}
 \end{equation}
 where $<n>$ denotes the average multiplicity of the produced particles.  
 Since eventually $\sigma_{tot} \sim 
ln^2 \left({s\over s_0}\right)$ the formula (\ref{n}) implies the power law 
increase 
of average multiplicity as $\left({s\over s_0}\right)^{\alpha_P(0)-1}$.  
It should be emphasized that the above statements hold at asymptotic energies 
and one can expect important finite energy corrections due to energy-momentum 
conservation in multiple inelastic collisions.  
These effects are automatically 
taken into account in various microscopic models of a pomeron
like, 
for instance, in the Dual Parton Model \cite{DTU}.\\

 The Regge phenomenology may also be applicable in the analysis of the deep 
inelastic scattering in the limit when the Bjorken variable $x$ is small.  
The inelastic lepton scattering (i.e. the reaction $l(p_l)+p(p) \rightarrow 
l^{\prime}(p_l^{\prime}) + anything$) is related through the one photon exchange 
approximation to the forward virtual Compton scattering $\gamma^*(Q^2)+
p(p) \rightarrow \gamma^*(Q^2)+p(p)$ where $Q^2=-q^2$, $q=p_l-p_l^{\prime}$ 
and $x=Q^2/2pq$. The small $x$ limit corresponds to $2pq>>Q^2$
i.e. 
to the high energy (Regge) limit of the forward virtual Compton
scattering.  The measured structure functions $F_2(x,Q^2)$ and 
$F_L(x,Q^2)$ are directly related to the total (virtual) photoproduction 
cross-sections $\sigma_T$ and $\sigma_L$ corresponding to transversely 
and longitudinally polarized photons: 
\begin{equation}
F_2(x,Q^2)={Q^2\over 4 \pi^2 \alpha}(\sigma_T +\sigma_L)
\label{fsigma2}
\end{equation}    
\begin{equation}
F_L(x,Q^2)={Q^2\over 4 \pi^2 \alpha} \sigma_L
\label{fsigmal}
\end{equation}
Assuming the conventional Regge pole parametrisation for $\sigma_{T,L}$ 
\begin{equation}
\sigma_{T,L}={4\pi^2 \alpha \over Q^2}\sum_i ({2pq\over Q^2})^{\alpha_i(0)-1}
C_{T,L}^i(Q^2)
\label{sregge}
\end{equation}
one gets  the following small $x$ behaviour for the structure functions: 
\begin{equation}   
F_{2,L}=\sum_i (x)^{1-\alpha_i(0)}C_{T,L}^i(Q^2)
\label{fregge}
\end{equation}
where the sum in (\ref{sregge},\ref{fregge}) extends over the pomeron and 
the reggeon contributions.  The experimental results from HERA show  that 
the structure function $F_{2}(x,Q^2)$ for moderate and large $Q^2$ values 
($Q^2 >$ 1.5 $GeV^2$ or so) grows more rapidly than expected on the basis of the 
straightforward extension of the Regge pole parametrization with the 
relatively small intercept of the effective pomeron ($\alpha_P(0) 
\approx$ 1.08) {\cite{THOMPSON,HALINA}. This result is consistent with  
 perturbative QCD which predicts much stronger increase of the parton 
 distributions and of the DIS structure functions with decreasing parameter 
 $x$ than that which would follow from equation (\ref{fregge}) with 
 $\alpha_P(0) \approx$ 1.08.  The high energy behaviour which follows 
 from perturbative QCD is often referred to as being related to 
 the "hard" 
 pomeron in contrast to the soft pomeron describing the high energy 
 behaviour of hadronic and photoproduction cross-sections. 
 The relevant framework for discussing the pomeron in
perturbative QCD and the small $x$ limit of parton 
distributions is the leading log$1/x$ (LL$1/x$) approximation which 
corresponds to the sum of those terms in the perturbative expansion 
where the powers of $\alpha_s$ are accompanied by the leading 
powers of ln($1/x$) \cite{BFKL,GLR,BCKK,ADM1,JK1}.  
At small $x$ the dominant role is played by the 
gluons and the quark (antiquark) distributions as well as the deep 
inelastic structure functions $F_{2,L}(x,Q^2)$ are also driven by the 
gluons through the $g \rightarrow q\bar q$ transitions.
Dominance of gluons in high energy scattering follows from the
fact that they carry spin equal to unity. 
 The basic  dynamical quantity at small $x$ is the   
unintegrated gluon distribution 
$f(x,Q_t^2)$ where now $x$ denotes the momentum fraction 
of a parent hadron carried by a gluon and $Q_t$  its transverse 
momentum.  The unintegrated distribution $f(x,Q_t^2)$ 
is related in the following way to the more familiar scale dependent 
gluon distribution $g(x,Q^2)$: 
\begin{equation}
xg(x,Q^2)=\int^{Q^2} {dQ_t^2\over Q_t^2} f(x,Q_t^2). 
\label{intg}
\end{equation}
In the leading $ln(1/x)$  approximation the unintegrated 
distribution $f(x,Q_t^2)$ satisfies 
the BFKL equation \cite{BFKL,GLR,BCKK,ADM1,JK1} which has the following form: 
$$
f(x,Q_t^2)=f^0(x,Q_t^2)+
$$
\begin{equation}
\bar \alpha_s \int_x^1{dx^{\prime}\over 
x^{\prime}} \int {d^2 q\over \pi q^2}
\left[{Q_t^2 \over (\mbox{\boldmath $q$}+
\mbox{\boldmath $Q_t$})^2} 
f(x^{\prime},(\mbox{\boldmath $q$}+
\mbox{\boldmath $Q_t$})^2)-f(x^{\prime},Q_t^2)\Theta(Q_t^2-q^2)\right]
\label{bfkl}
\end{equation}
where 
\begin{equation}
\bar \alpha_s={3\alpha_s\over \pi}
\label{alphab}
\end{equation}
This equation sums the ladder diagrams with gluon exchange accompanied 
by virtual corrections which are responsible for the gluon reggeization. 
The first and the second 
terms  on the right hand side of  eq. (\ref{bfkl}) correspond 
to  real gluon emission with $q$ being the transverse 
momentum of the emitted gluon, and to the virtual corrections respectively. 
$f^0(x,Q_t^2)$ is a suitably defined inhomogeneous term. 
For the fixed coupling case  eq. (\ref{bfkl}) can be solved 
analytically and the leading behaviour of its solution 
at small $x$ is given by the 
following expression:
\begin{equation} 
f(x,Q_t^2) \sim (Q_t^2)^{{1\over 2}} {x^{-\lambda_{BFKL}}\over 
\sqrt{ln({1\over x})}} exp\left(-{ln^2(Q_t^2/\bar Q^2)\over 2 \lambda^"
ln(1/x)} \right)
\label{bfkls}
\end{equation} 
with 
\begin{equation}
\lambda_{BFKL}=4 ln(2) \bar \alpha_s
\label{pombfkl}
\end{equation} 
\begin{equation} 
\lambda^"=\bar \alpha_s 28 \zeta(3) 
\label{diff}
\end{equation}
where the Riemann zeta function $\zeta(3) \approx 1.202$.  The 
parameter $\bar Q$ is of nonperturbative origin. 
The quantity $1+ \lambda_{BFKL}$ is equal to the intercept of the so -  
called BFKL pomeron. Its potentially large magnitude ($\sim 1.5$) 
should be contrasted with the intercept $\alpha_{soft} \approx 1.08$ 
of the (effective) "soft" pomeron which has been determined 
from the phenomenological analysis of the high energy behaviour 
of hadronic and photoproduction total cross-sections \cite{DOLACX}. 
The solution of the BFKL equation 
reflects its diffusion pattern which  
is the direct consequence of the absence of transverse momentum ordering 
along the gluon chain.   
The interrelation between the diffusion of transverse momenta towards 
both the infrared and ultraviolet regions {\bf and} the increase of gluon 
distributions 
with decreasing $x$ is a  characteristic property of QCD at low $x$.  
It has important consequences for the structure of the hadronic final 
state in deep inelastic scattering at small $x$ 
\cite{JK1}.\\

In practice one introduces the running coupling $\bar \alpha_s(Q_t^2)$ 
in the BFKL equation (\ref{bfkl}). This requires the introduction of an 
infrared 
cut-off to prevent entering the infrared region where the 
coupling becomes large. The effective intercept $\lambda_{BFKL}$ 
found by numerically solving the equation depends   
on the magnitude of this cut-off. The running coupling does 
also affect the diffusion pattern of the solution. 
 The effective intercept $\lambda_{BFKL}$ turns
 out 
to be also sensitive 
on the (formally non-leading) additional constraint $q^2 < Q_t^2x^{\prime}/x$ 
in the real emission term in eq. (\ref{bfkl}) which 
follows from the requirement that the virtuality of the 
last gluon in the chain is dominated by $Q_t^2$ \cite{BO,KMSGLU}.
The impact of the momentum 
cut-offs on the solution of the BFKL equation has also been discussed 
in refs. \cite{PVLC,MCDG}. In  impact parameter representation 
the BFKL equation offers an 
interesting  
interpretation in terms of colour dipoles \cite{DIPOLE}. It 
should also be emphasised that the complete calculation of the next-to-leading 
corrections to the BFKL equation has recently become presented in ref.
 \cite{NLX}.\\

The structure functions $F_{2,L}(x,Q^2)$ are driven  at small $x$ 
by the gluons 
 and are related in the following way to the unintegrated distribution $f$: 
\begin{equation}
F_{2,L}(x,Q^2)=\int_x^1{dx^{\prime}\over x^{\prime}}\int 
{dQ_t^2\over Q_t^2}F^{box}_{2,L}(
x^{\prime},Q_t^2,Q^2)f({x\over x^{\prime}},Q_t^2). 
\label{ktfac}
\end{equation}
The functions  $F^{box}_{2,L}(x^{\prime},Q_t^2,Q^2)$ may be regarded as  the 
structure 
functions of the off-shell gluons with  virtuality  
$Q_t^2$.  
They are described by the quark box (and crossed box) diagram contributions 
to the 
photon-gluon interaction.   
The small $x$ behaviour of the structure functions reflects the small 
$z$ ($z = x/x^{\prime}$) behaviour of the gluon distribution $f(z,Q_t^2)$.\\

Equation (\ref{ktfac}) is an example of the "$k_t$ factorization theorem" 
which relates measurable quantities (like DIS structure functions) to 
the convolution in both longitudinal as well as in transverse momenta of the 
universal gluon distribution $f(z,Q_t^2)$ with the cross-section 
(or structure function) describing the interaction of the "off-shell" gluon 
with the hard probe \cite{KTFAC,CIAFKT,MK,CIAFQQ,CATHEP}.  
The $k_t$ factorization theorem is the basic tool for 
calculating the observable quantities in the small $x$ region in terms of the 
(unintegrated) gluon distribution $f$ which is the solution of the BFKL 
equation.\\ 

The leading - twist part of the $k_t$ factorization formula can be rewritten 
in a collinear factorization form.  The leading small $x$ effects are then 
automatically resummed in the  
 splitting functions and in the coefficient functions. The $k_t$ 
factorization theorem   can in fact be used as the tool for calculating 
these quantities.   
The small $x$ resummation effects within the conventional QCD evolution 
formalism have recently been discussed in refs. \cite{EKL,HBRW,BFORTE,FRT}. 
  One finds in general that at the moderately small 
values of $x$ which are relevant for 
the HERA measurements,   the small $x$ resummation effects in the 
splitting function $P_{qg}$ have a much stronger impact on $F_{2}$ than 
the small $x$ resummation in the splitting function $P_{gg}$. 
It should also be remembered that the BFKL effects 
in the splitting function $P_{qg}(z,\alpha_s)$ can significantly affect 
extraction of the gluon distribution out of the experimental data on the 
slope of the structure function $F_2(x,Q^2)$ which is based on the following relation:
\begin{equation}
Q^2{\partial F_2(x,Q^2)\over \partial Q^2} \simeq 
2 \sum_i e_i^2 \int_x^1 dz P_{qg}(z,\alpha_s(Q^2)){x\over z}
g({x\over z},Q^2)
\label{slopef2}
\end{equation}
 
A more general treatment of the gluon ladder than that which follows 
from the BFKL formalism is  provided by 
the Catani, Ciafaloni, Fiorani, Marchesini (CCFM) equation based on
 angular ordering along the gluon chain 
\cite{CIAF,CCFM,KMS1}.  
This equation embodies both the BFKL equation at small $x$ and the 
conventional Altarelli-Parisi evolution at large $x$.  
The unintegrated gluon distribution $f$ now acquires   
dependence upon an additional scale $Q$ 
which specifies the maximal angle of gluon 
emission.  
The CCFM equation has the following form : 
$$
f(x,Q_t^2,Q^2)=\hat f^0(x,Q_t^2,Q^2)+ $$
\begin{equation}
\bar \alpha_s
\int_x^1{dx^{\prime}\over 
x^{\prime}} \int {d^2 q\over \pi q^2} \Theta
(Q-qx/x^{\prime})\Delta_R({x\over x^{\prime}},Q_t^2,q^2)
{Q_t^2 \over (\mbox{\boldmath $q$}+
\mbox{\boldmath $Q_t$})^2} 
f(x^{\prime},(\mbox{\boldmath $q$}+
\mbox{\boldmath $Q_t$})^2,q^2))      
\label{ccfm}
\end{equation}
where the theta function $\Theta(Q-qx/x^{\prime})$ reflects the angular 
ordering constraint on the emitted gluon.  
The "non-Sudakov" form-factor $\Delta_R
(z,Q_t^2,q^2  )$ is now given by the following formula: 
\begin{equation}
\Delta_R(z,Q_t^2,q^2)=exp\left[-\bar \alpha_s\int_z^1 {dz^{\prime}
\over z^{\prime}} \int {dq^{\prime 2}
\over q^{\prime 2}}\Theta (q^{\prime 2}-(qz^{\prime})^2)
\Theta (Q_t^2-q^{\prime 2})\right]
\label{ns}
\end{equation}
Eq.(\ref{ccfm}) still contains only the singular term of the 
$g \rightarrow gg$ splitting function  
at small $z$. Its generalization which would  
include 
remaining parts of this vertex (as well as quarks) is possible.  
The numerical analysis of this equation was presented in ref. \cite{KMS1} 
. \\

The HERA data can be described quite well using the BFKL and CCFM equations 
combined with the factorization formula (\ref{ktfac})  
\cite{KMSGLU,CCFMF2}. One can however 
obtain satisfactory description of the HERA data staying 
within the scheme   
 based on the Altarelli-Parisi equations alone 
without the small $x$ resummation effects being included in the formalism 
\cite{MRS,GRV}.   
In the latter case the singular small $x$ behaviour of the gluon 
and sea quark distributions  
 has to be introduced in the parametrization of the starting 
distributions at the moderately large reference scale $Q^2=Q_0^2$   
 (i.e. $Q_0^2 \approx 4 GeV^2$ or so) \cite{MRS}.  One can also 
generate steep behaviour dynamically starting from  
non-singular "valence-like" parton distributions at some very low 
scale $Q_0^2=0.35GeV^2$ \cite{GRV}. In the latter case the gluon and sea 
quark 
distributions exhibit  "double logarithmic behaviour" \cite{DL} 
\begin{equation}
F_2(x,Q^2) \sim exp \left(2\sqrt{\xi(Q^2,Q_0^2)ln(1/x)}\right)
\label{dlog}
\end{equation}
where 
\begin{equation}
\xi(Q^2,Q_0^2)=\int_{Q_0^2}^{Q^2}{dq^2\over q^2}{3\alpha_s(q^2)\over \pi} . 
\label{evlength}
\end{equation} 
For very small values of the scale $Q_0^2$ the evolution length $\xi(Q^2,
Q_0^2)$  
can become large for moderate and large values of $Q^2$ and the "double 
logarithmic" behaviour (\ref{dlog}) is, within the limited region of $x$,  
similar to that corresponding to the power like increase of the type 
$x^{-\lambda}$, $\lambda \approx 0.3$.\\

It is expected that absence of transverse momentum ordering along the gluon 
chain which leads to the correlation between the increase of the 
structure function  with  decreasing $x$ and the diffusion 
of transverse momentum should reflect itself in the behaviour of 
less inclusive 
quantities than the structure function $F_2(x,Q^2)$.  The dedicated 
measurements of the low $x$ physics which are particularly sensitive 
to this correlation are the deep inelastic plus 
jet events, transverse energy flow 
in deep inelastic scattering, production of jets separated by the large 
rapidity gap and dijet production in deep inelastic scattering.  \\

In principle  deep inelastic lepton scattering containing a measured jet 
can provide a very clear test of the BFKL dynamics at low $x$
\cite{MJET,BJET,WKJET,KMSJET,BJET1}.  
The idea is to study deep inelastic ($x,Q^2$) events which 
contain an identified jet ($x_j,k_{Tj}^2$) where 
$x<<x_j$ and $Q^2 \approx k_{Tj}^2$.  Since we choose events with 
$Q^2 \approx k_{Tj}^2$ the leading order QCD evolution  (from $k_{Tj}^2$ to 
$Q^2$) is neutralized and attention is focussed on the small $x$, or rather 
small $x/x_j$ behaviour.  The small $x/x_j$ behaviour
of jet production is generated  by the gluon radiation. Choosing the 
configuration $Q^2 \approx k_{Tj}^2$ we eliminate by definition  
gluon emission which corresponds to strongly ordered transverse momenta 
i.e.  that emission which is responsible for the LO QCD evolution.  
The measurement of jet production in this configuration may therefore test 
more directly the $(x/x_j)^{-\lambda}$ behaviour which is generated 
by the BFKL equation where the transverse momenta are not ordered.  
The recent H1 results concerning  deep inelastic plus jest events 
are consistent with the increase of the cross-section with decreasing 
$x$ as predicted by the BFKL dynamics {\cite{BJET1,EWELINA}. \\

Complementary measurement to deep inelastic plus forward jet is the 
deep inelastic scattering accompanied by the forward prompt photon or 
forward prompt $\pi^0$. \cite{KLM1,KLM2}\\ 

Conceptually similar process is that of the two-jet production 
separated by a large rapidity gap $\Delta y$ in hadronic 
collisions or in photoproduction
\cite{DDUCA,JAMESJ}.  
Besides the characteristic $exp(\lambda \Delta y)$ dependence 
of the two-jet cross-section one expects significant 
weakening of the azimuthal back-to-back correlations 
of the two jets.  This is the direct consequence of the 
absence of transverse momentum ordering along the gluon  
chain.\\

Another measurement which should be sensitive to the QCD 
dynamics at small $x$ is that of the transverse energy flow in deep inelastic 
lepton scattering in the central region away from the 
current jet and from the proton remnant  
\cite{KMSET}.  
The  BFKL dynamics predicts 
in this case a substantial amount of transverse energy  which should 
increase with decreasing $x$.  The experimental data 
are consistent with this theoretical expectation \cite{KUHLEN}.    
\\
Absence of transverse momentum ordering  also implies weakening of the 
back-to-back azimuthal correlation of dijets produced close 
to the photon fragmentation region \cite{DICKJET,AGKM}. \\

Another important process which is sensitive to the small $x$ dynamics 
is  the deep inelastic diffraction \cite{ZDIF,H1DIF}. 
Deep inelastic diffraction in $ep$ inelastic scattering is a process: 
\begin{equation}
e(p_e)+p(p) \rightarrow e^{\prime}(p_e^{\prime}) + X +p^{\prime}(p^{\prime})
\label{disdif}
\end{equation}
where there is a large rapidity gap between the recoil proton 
(or excited proton) and the hadronic system $X$.  
To be precise   
process (\ref{disdif}) reflects the diffractive disssociation 
of the virtual photon.  Diffractive dissociation is described by the following 
kinematical variables: 
\begin{equation}
\beta={Q^2\over 2 (p-p^{\prime})q}
\end{equation}
\begin{equation}
x_P={x\over \beta}
\end{equation}
 \begin{equation}
t= (p-p^{\prime})^2. 
\label{difv}
\end{equation}
Assuming that  diffraction dissociation is dominated by the pomeron 
exchange  and that the pomeron is described by 
a Regge pole one gets the following factorizable expression for the 
diffractive structure function \cite{DOLADIF,ORSAYDIF,CTEQD,JAMESD,KGBK}:
\begin{equation}
{\partial F_2^{diff}\over \partial x_P \partial t}= f(x_P,t)F_2^P(\beta,Q^2,t)
\label{difsf}
\end{equation}
where the "flux factor" $f(x_P,t)$ is given by the following formula
:
\begin{equation}
f(x_P,t)=N{B^2(t)\over 16\pi} x_P^{1-2\alpha_P(t)}
\label{flux}
\end{equation}
with $B(t)$ describing the pomeron coupling to a proton and $N$ being the 
normalisation factor. Equation (\ref{difsf}) for the diffractive
structure function follows from equation  (\ref{difdisab}) for
the diffractive cross-section  $\gamma^* + p \rightarrow X
+p$.  The function $F_2^P(\beta,Q^2,t)$
is the pomeron structure function which in the (QCD improved) parton model 
is related in a standard way to the quark and antiquark distribution 
functions in a pomeron.  
\begin{equation}
F_2^P(\beta,Q^2,t)=\beta \sum e_i^2[q_i^P(\beta,Q^2,t)+ \bar 
q_i^P(\beta,Q^2,t)]
\label{f2pom}
\end{equation} 
with $q_i^P(\beta,Q^2,t)=\bar q_i^P(\beta,Q^2,t)$.  The variable 
$\beta$ which is the Bjorken scaling variable appropriate for 
deep inelastic lepton-pomeron "scattering",  has the meaning of the 
momentum fraction of the pomeron carried by the 
probed quark (antiquark).  The quark distributions in a pomeron 
are assummed to obey the standard Altarelli-Parisi evolution equations: 
\begin{equation}
Q^2{\partial q^P\over \partial Q^2}=P_{qq} \otimes q^P + P_{qg} \otimes g^P
\label{app}
\end{equation}
with a similar equation for the evolution of the gluon distribution 
in a pomeron.  The first 
term on the right hand side of the eq. (\ref{app}) becomes negative 
at large $\beta$,  while the second term remains positive and 
is usually very small at large $\beta$ unless the gluon distributions 
are large and have a hard spectrum.\\

 The data suggest that the slope 
of $F_2^P$ as the function of $Q^2$ does not change sign even 
at relatively large values of $\beta$.  This favours the hard 
gluon spectrum in a pomeron \cite{CAPHG,KGBJP}, and   
should be contrasted with the behaviour of the structure function 
of the proton which, at large $x$, decreases with increasing $Q^2$.   
The data on inclusive diffractive production  favour the soft pomeron 
with relatively low intercept.  
 The diffractive production of vector mesons 
 seems to require a "hard" pomeron contribution 
\cite{THOMPSON,HALINA} . It has also 
been pointed out  that the factorization property 
(\ref{difsf}) may 
not hold in models based 
entirely on perturbative QCD when the pomeron is represented 
by the BFKL ladder \cite{KOLYA,BARTELSD}.  The factorization does not also 
hold when the exchange of the secondary Regge poles besides pomeron 
becomes important \cite{KGBSL,PWAW}.  The contribution of secondary Reggeons 
is expected to be significant at moderately small $x_P$ and small values of 
$\beta$.  The recent HERA results show violation of factorization in this 
region \cite{PWAW}. Finally let us point out that 
there exist also models of deep inelastic diffraction which do not 
rely on the pomeron exchange picture \cite{BUCHMD,EIR}.\\
  
To summarize we have recalled in this talk the Regge phenomenology 
of diffractive scattering based on the effective "soft" pomeron 
exchange and have  briefly described the QCD expectations 
for deep inelastic lepton scattering at low $x$.  
Perturbative QCD predicts  indefinite increase of gluon 
distributions 
with decreasing $x$ which generates similar increase of the structure 
functions through the 
$g \rightarrow q \bar q$ transitions. 
The indefinite growth of parton distributions cannot go on forever 
and has to be eventually stopped by parton screening which leads 
to the parton saturation.  Most probably however 
the saturation limit is still irrelevant for the small $x$ region 
which is now being probed at HERA. Besides discussing the theoretical and 
phenomenological issues related to the description of the structure 
function $F_2$ at low $x$ we have also emphasised 
the role of studying the hadronic final state in deep inelastic scattering 
for probing the QCD pomeron. Finally let us point out that the  recent experiments at HERA  cover very 
broad range of $Q^2$ including the region of low and moderately 
large values of $Q^2$.  Analysis of the structure functions in
this  transition region is very interesting \cite{BBJK}               
 and may help to understand 
possible relation (if any) between the soft and hard pomerons.  
\section*{Acknowledgments}
I thank Jorge Dias de Deus for his very kind invitation to 
the Multiparticle Dynamics Symposium, for very warm hospitality in 
Faro and for organizing an excellent meeting.  
I thank Barbara Bade\l{}ek, Krzysztof Golec-Biernat, 
Alan Martin and Peter Sutton for most enjoyable research collaborations  
on some of the problems presented in this review.     
This research has been supported in part by 
 the Polish State Committee for Scientific Research grant 
 2 P03B 231 08 and 
the EU under contracts Nos. CHRX-CT92-0004/CT93-357.\\

\end{document}